\documentstyle[titlepage,12pt]{article}
\newcommand{\be}{\begin{equation}}
\newcommand{\ee}{\end{equation}}
\newcommand{\bear}{\begin{eqnarray}}
\newcommand{\ear}{\end{eqnarray}}

\begin{document}
\begin{titlepage}
\begin{flushright}
HD-THEP-96-03\\
PITHA 96/03
\end{flushright}
\vspace{0.8cm}
\begin{center}
{\bf\LARGE Some Remarks on the Search\\}
{\bf\LARGE for CP Violation in Z-Decays
\footnote{Research supported  by BMBF, contract No. 05 6HD 93P(6)}\\}
\vspace{2cm}
\centerline{ \bf W. Bernreuther$^a$ and O. Nachtmann$^b$}

\vspace{1cm}
\centerline{$^a$Institut f. Theoretische Physik,
RWTH Aachen, D-52056 Aachen, Germany}
\centerline{$^b$Institut f. Theoretische Physik,
Universit\"at Heidelberg, D-69120 Heidelberg, Germany}

\vspace{3cm}

{\bf Abstract:}\\
\parbox[t]{\textwidth}
{We discuss some issues arising in the search for CP violation in
the decays $Z\to\tau^+\tau^-$ and $Z\to b\bar bX$. The form-factor
and the effective Lagrangian approach to parametrize CP-violating
effects are compared. We emphasize the interest to study
both real and imaginary parts of CP-violating form factors like
the weak dipole moment form factor $d_\tau^Z(m^2_Z)$ of the
$\tau$-lepton. We propose an ``optimal'' way to search for CP
violation when $Z\to b\bar bX$ events have to be selected on a
statistical basis  from the total number of $Z$ hadronic decays.}
\end{center}
\end{titlepage}
\newpage

\section{Introduction}
The anomalous magnetic dipole moments (AMM)
of the electron and muon have played
an important role in precision tests of quantum electrodynamics.
Future measurements of the muon
AMM may  eventually become sensitive to quantum corrections as
predicted by the Standard Model (SM) \cite{1}.
Searches for non-zero electric dipole moments (EDM), for instance of
the electron or the neutron,
are motivated by the sensitivity of these moments
to CP-violating interactions
beyond the one parametrized by the phase in the
Kobayashi-Maskawa matrix of the SM. In precise terms these dipole
moments are the chirality-flipping form factors in the fermion photon
vertex at zero momentum transfer. The fermion
Z boson vertex contains analogous
chirality-flipping form factors (see below).
In renormalizable theories these
moments cannot appear as elementary couplings but  are
induced by quantum corrections. It can be argued that
the tau dipole form factors are probably  much more sensitive to new
physics than
the corresponding moments of the electron or muon. For instance,
diagrams with exchange of
a neutral Higgs boson  in the gauge boson fermion vertex induce
contributions to the dipole form factors which
scale with the third power of
the external fermion mass.
It is therefore of great interest to determine the tau moments
as precisely as possible  \cite{2}-\cite{8}.

Most direct or indirect experimental determinations
which were made so far
for the tau moments
involve time like momentum transfers.
In particular the most precise direct determination todate is
that of the weak dipole
moment (WDM) form factor of the tau at $q^2=m^2_Z$ in ref. \cite{9}
which combines the results of
the OPAL \cite{10}, \cite{11} and ALEPH collaborations
\cite{12}, \cite{13}. The DELPHI collaboration
\cite{14}
obtained upper limits on the AMM and
EDM form factors of the tau lepton at $q^2=0$ from an analysis of
 $e^+e^-\to \tau^+\tau^-\gamma$ events.

For $q^2=m_Z^2$
the form factors can be \underbar{complex}. On the other hand,
effects of new physics beyond the SM, in particular
of new CP-violating
interactions with an intrinsic energy scale $\Lambda\gg m_Z$ can
quite generally be parametrized by an effective Lagrangian ${\cal L}
_{\rm eff}$. This Lagrangian can be expanded in terms of hermitian operators
of dimensions 5,6,... constructed from SM fields. The coupling
constants multiplying the hermitian operators in ${\cal L}_{\rm eff}$
must be \underbar{real}. How do these \underbar{real} coupling
constants relate to the \underbar{complex} form factors above?

In this note we recall for the convenience
of the experimentalists the relation between the coupling constants
of the effective Lagrangian and the form factors. We discuss
the definitions and a few properties of
these form factors.
Most experiments used the observable $T_{33}$ \cite{3,5} which determines
the real
part of the weak dipole moment form factor.
But we emphasize that experiments which are sensitive to form factors
at time like momentum transfers  should measure both their real and
imaginary parts, as was done for the WDM form factor of the tau lepton in
{\cite{11}}. Furthermore, we make some remarks concerning
an ``optimal'' strategy for searching for
CP violation when one has to select a class of events
on a statistical basis. As an example we discuss the CP violation
search in $Z\to b\bar b X$ (cf. \cite{15}).

\section{Form Factors}

Precision measurements of the electromagnetic moments of the
electron and muon amount to measuring the form factors of the respective
electromagnetic current matrix element at zero momentum transfer.
Using Lorentz covariance and current conservation, the matrix element
of the electromagnetic current for a lepton $f=e,\mu,\tau$
can be written as follows (for a discussion of form factors and CP
violation cf. e.g. \cite{16}):

\begin{eqnarray}\label{eff1}
&&<f(p')|eJ^{em}_{\mu}(0)|f(p)>  =  -e{\bar u}(p')\Gamma_{\mu}u(p),
\nonumber\\
&&(f=e,\mu,\tau).\end{eqnarray}
Here $eJ^{em}_\mu$ is the electromagnetic current operator,
$e$ is the positron charge $(e>0)$ and the vertex factor
$\Gamma_\mu$ can be decomposed as:
\begin{equation}\label{eff2}
\Gamma_{\mu}  =  F_1\gamma_{\mu} + F_2 i\sigma_{\mu\nu}q^{\nu}/2m_f
+F_3 \sigma_{\mu\nu}\gamma_5 q^{\nu}/2m_f + F_4(\gamma_{\mu}\gamma_5
q^2 - 2m_f \gamma_5 q_{\mu}),
\end{equation}
where $m_f$ is the mass
of the fermion $f$, $q = p'- p$,  and  $F_i = F_i(q^2)$.
Our metric and $\gamma$-matrix conventions are as in \cite{161}.
The static quantites $F_i(0) (i=1,2,3)$ which are measured in the soft
photon limit $q\to 0$ are real and $SU(2)\times U(1)$
gauge-invariant.
They are defineable as the residues of
the photon pole $\propto 1/q^2$ in physical scattering
amplitudes.  Recall that by charge renormalization $F_1(0)=1$.
The magnetic moment $\mu_f$, the AMM $a_f^\gamma$, and the
P- and T-violating EDM $d^\gamma_f$ are given by
\begin{eqnarray}\label{3}
\mu_f&=&-\frac{e}{2m_f}[F_1(0)+F_2(0)],\nonumber\\
a^\gamma_f&=&F_2(0),\nonumber\\
d^\gamma_f&=&\frac{e}{2m_f}F_3(0).\end{eqnarray}
The form factor $F_4(0)$ is related to the ``anapole moment''
of the fermion $f$ \cite{17}. The anapole moment operator $\underline
{\vec{\cal A}}$ is defined by
\be\label{4}
\underline{\vec{\cal A}}(t)=-e\pi\int d^3x
|{\vec x}|^2\vec J^{em}(\vec x,t),\ee
and the anapole moment ${\cal A}_f$ of the fermion $f$ by
\be\label{5}
<f(p',s')|\underline{\vec{\cal A}}(0)|f(p,s)>\Bigl|_{p'=p=p_R}
={\cal A}_f<f(p',s')|\vec S|f(p,s)>\Bigr|_{p'=p=p_R}\ee
Here $\vec S={\vec \sigma}/2$ denotes the spin operator and
$p_R$ the 4-momentum
vector of the fermion $f$ at rest.
A simple calculation gives
\be\label{6}
{\cal A}_f =8\pi eF_4(0).\ee
The anapole moment is a P-odd and T-even quantity. When defined as
in (\ref{6}), it is invariant
under electromagnetic gauge transformations but not under the full
$SU(2)\times U(1)$ gauge transformations of the SM. (Its
SM value  was computed in
\cite{Musolf} in the general $R_\xi$ gauge.) This can be
seen as follows. The form factor $F_4$ in (\ref{eff2}) multiplies
a vector structure $\gamma_\mu\gamma_5 q^2-2m_fq_\mu\gamma_5$ which
does \underbar{not} lead to a pole term in $q^2$ from photon
exchange in physical amplitudes. Indeed, multiplying with the
photon propagator $\propto 1/q^2$ we get
\[\gamma^\mu\gamma_5-2m_f\frac{q^\mu}{q^2}\gamma_5.\]
The first term has no pole and the term
$q^\mu/q^2$ gives a
vanishing contribution due to current conservation when contracted
with the vertex on the other side of the propagator line (Fig. 1).
Thus the anapole moment corresponds to a contact interaction
at small $|q^2|$ and its value is only fixed relative to the convention
on $\gamma-Z$ mixing in their $2\times 2$ propagator matrix.

In \cite{Stuart} the anapole moment is defined as being the axial vector
contact interaction with an external electromagnetic current.
This definition includes corrections from both photon and Z boson exchange
and leads to an expression which is gauge-invariant with respect to the
full electroweak theory.

For $q^2\not=0$ the form factors $F_i(q^2)$ have infrared
divergences due to ordinary QED radiative corrections. We will
ignore these infrared divergences for the moment.

Tau pair production in $e^+ e^-$ collisions
is described by the corresponding S matrix element. It involves
in particular the vertex
 $\gamma^*(q) \to {\bar f}(p)+f(p')$
where now $q = p'+ p$ and the vertex where the $\gamma^*$ is
replaced by a $Z$. The $\gamma^*$ vertex
is obtained from (\ref{eff1}),(\ref{eff2}) by crossing. The
form factors
$F_i(q^2)$ develop imaginary parts for $q^2 > s_0$. The threshold
value $s_0$ may depend on the form factor. Usually one has normal
thresholds in which case
$s_0=(m_1+m_2)^2$ where $m_1,m_2$ are the masses of
the lightest pair of intermediate particles, which can couple
to $f\bar f$ and the current in question.
${\rm Re} F_i$ and ${\rm Im} F_i$ are
related by a dispersion integral. For $q^2
\neq 0$ the form factors are in general not
invariant with respect to the gauge fixing
conventions of the electroweak theory.

Gauge-invariant expressions at the Z resonance are obtained
from a Laurent expansion of the S matrix element at its
Z pole in the complex $s$ plane. For the discussion of CP violation
effects at the Z resonance it is legitimate to neglect the non-resonant
terms and to study the residue of the pole in a weak coupling expansion,
typically to one-loop order. Then
 the S matrix element for the on-shell process
$Z\to \tau^+\tau^-$ applies. It can be decomposed as follows:
\be\label{7}
<\tau^-(p_1)\tau^+(p_2)|S|Z(p,\epsilon)> = i
(2\pi)^4\delta(p-p_1-p_2)e \epsilon^{\mu}{\bar
u}(p_1)\Lambda_{\mu}v(p_2),\ee
where
\be\label{8}
\Lambda_{\mu} =  G_1(m^2_Z)
\gamma_{\mu} + G_2 (m^2_Z)i\sigma_{\mu\nu}p^{\nu}/2m_{\tau}
+G_3 (m^2_Z)\sigma_{\mu\nu}\gamma_5 p^{\nu}/2m_{\tau} +
G_4(m^2_Z)\gamma_{\mu}\gamma_5,
\ee
and $p = p_1+p_2$. Note the following: \\
1) The vertex function (\ref{8}) is electroweak gauge-invariant
 if all the particles are on-shell.\\
2) The vertex function for the coupling
of $f\bar f$ to an off-shell $Z$ boson with
invariant mass squared $p^2=s$ is again
given by (\ref{8}) but with $G_i(m_Z^2)
\to G_i(s)$. Also the term $iG_5(s)p_{\mu}
 + G_6(s)
p_{\mu}\gamma_5$ containing two more form factors
has to be added in (\ref{8}). However, the form factors $G_{5,6}$
do not contribute to $e^+e^-\to \tau^+\tau^-$
in the
limit of vanishing electron mass.\\
3) The form factors $G_i(p^2=m^2_Z)$ are in general complex.
The form factors $G_3$ and $G_5$ are CP-violating.\\
4) For a check of the time reversal symmetry T the \underbar{decay}
reaction $Z\to f\bar f$ has to be compared with the
\underbar{production} reaction $f\bar f\to Z$. The form factors for
the latter reaction can be obtained from (\ref{8}) assuming CPT
invariance or crossing symmetry, which are, of course, both valid
in a local relativistic quantum field theory (cf. e.g. \cite{171}). Thus,
the reaction $Z\to f\bar f$ by itself allows straightforward CP-tests,
whereas for T-tests one has to assume CPT invariance and then
T invariance is equivalent to CP invariance.\\
5) In analogy to the electromagnetic moments at $q^2=0$ we can define
electromagnetic and weak anomalous magnetic moment (AMM,WMM)
as well as electric and weak dipole moment (EDM, WDM) form factors by
\bear\label{9}
a^\gamma_\tau(s)&=&F_2(s),\nonumber\\
a^Z_{\tau}(s) & = & G_2(s),\ear
\bear\label{10}
d^\gamma_\tau(s)&=&\frac{eF_3(s)}{2m_\tau},\nonumber\\
d^Z_{\tau}(s) & = & \frac{e G_3(s)}{2m_\tau}.\ear
The SM value of $a^Z_{\tau}(m^2_Z)$, which is complex,  was computed in
\cite{8}.
The CP-violating form factors $d^{\gamma,Z}_{\tau}(s)$
are unmeasurably small within the SM, but they are generated to one-loop
order in a number of SM extensions. The branch point at which these
form factors develop an imaginary part depends on the model
(cf. \cite{18,19}
and below). As an example of a SM extension we mention models
with neutral Higgs bosons $\varphi$ which couple both to
scalar and pseudoscalar quark and lepton currents ( cf., for instance
\cite{18}). Neglecting the electron mass then to one-loop approximation
$\varphi$ exchange induces the CP-violating contributions
to the S matrix element of $e^+e^-\to\tau^+\tau^-$  depicted in Fig.2.
These contributions lead to
form factors $d^\gamma_\tau(s)$ and $d^Z_{\tau}(s)$. Clearly they have imaginary
parts for $s > 4 m^2_{\tau}$. The contribution Fig. 2b  to the WDM form factor
has its branch point at $(m_Z+m_{\varphi})^2$.

\section{Effective Lagrangian versus Form Factors}

Let us discuss now in general terms interactions beyond the SM with an
intrinsic mass scale $\Lambda$ much larger than the
characteristic energy scale of a reaction, which in our case is $m_Z$:
\be\label{11}  \Lambda \gg m_Z.\ee
Then the effects of these new interactions at or below the Z scale
can be parametrized by an
effective Lagrangian which consists of a sum of hermitian \underbar{local}
operators containing only
SM fields with real coupling constants. The operators and their couplings
in  ${\cal L}_{\rm eff}$
can be classified according to their
(mass) dimensions.
\underbar {After} spontaneous symmetry breaking the effective Lagrangian for
CP violation in the $\gamma$ fermion  and $Z$ fermion systems reads,
including operators up to dimension five \cite{2,20}:
\be\label{12}
{\cal L}_{{\rm eff},1}=-\frac{i}{2}\bar f\sigma_{\mu\nu}\gamma_5f
\left[\tilde d_f^{\gamma(0)}F^{\mu\nu}+
\tilde d_f^{Z(0)}Z^{\mu\nu}\right] - \tilde a_f^{(0)}
(\bar f i\gamma_5f) Z_{\mu} Z^{\mu},
\ee
where $F^{\mu\nu}=\partial^\mu A^\nu-\partial^\nu A^\mu$ is
the electromagnetic field strength tensor and
$Z^{\mu\nu}=\partial^\mu Z^\nu-\partial^\nu Z^\mu$.
The constants $\tilde d_f^{\gamma(0)},\tilde d_f^{Z(0)},
\tilde a_f^{(0)}$ in
(\ref{12}) are real. If we calculate the dipole form factors
(\ref{10}), which follow from (\ref{12}), to
\underbar{ zeroth order} in the SM
coupling constants, we get:
\bear\label{14}
d_f^\gamma(s)&=&\tilde d_f^{\gamma(0)},\nonumber\\
d_f^Z(s)&=&\tilde d_f^{Z(0)}.\ear
(In fact, the electromagnetic coupling $e$ is contained in
$\tilde d^{\gamma,Z(0)}$ which  we do not count.)
This implies in particular that the
imaginary parts of the form factors (\ref{10}) vanish at zeroth
order. The relations (\ref{14}) are modified when SM corrections
are taken into account. To first and second order in the SM couplings
there are 2 types of such corrections:\\
1) Corrections which are of first order both in the SM  and
in the CP-violating couplings which differ from the dipole couplings. Typical
diagrams for this type of corrections are shown in Fig. 3a which
involves the $ZZ\tau\tau$ interaction of (\ref{12}), and in Fig. 3b
 which involves the couplings (cf. \cite{4})
\bear\label{15}
{\cal L}_{{\rm eff},2}&=&\sum_f\Bigl\{\tilde h_{1f}^{(0)}(\bar f f)(\bar
\tau i\gamma_5\tau)\nonumber\\
&&+\tilde h_{2f}^{(0)}(\bar f i\gamma_5 f)(\bar\tau \tau)\nonumber\\
&&+\tilde h^{(0)}_{3f}\epsilon^{\mu\nu\rho\sigma}
(\bar f\sigma_{\mu\nu}f)(\bar\tau\sigma_{\rho\sigma}\tau)
\Bigr\}.\ear
Here $f$ can be any of the fundamental fermions of the SM.
Actually, the scalar-pseudoscalar  interactions in (\ref{15})
generate EDM and WDM form factors of the $\tau$
through  diagram Fig. 3b
only if $f=\tau$.
For $f\neq\tau$ the contribution of the
first term in (\ref{15}) to the diagram Fig. 3b
is zero, whereas the second term in (\ref{15}) leads to the scalar
form factor $G_5$ mentioned in sect. 2. For $f=\tau$ one can use the
Fierz identity $(\bar\tau \tau)(\bar\tau\gamma_5\tau)
= -(\bar\tau\sigma_{\mu\nu}\tau)(\bar\tau\sigma^{\mu\nu}\gamma_5\tau)/12$.
That is, in this case only the coupling $\tilde h^{(0)}_{3\tau}$
appears in (\ref{15}).  Fig. 3b and the analogous diagram with
an external photon lead to complex WDM and EDM form factors
above the $\tau^+\tau^-$ threshold.\\
2) Corrections involving the dipole couplings of (\ref{12}).
Typical diagrams for this type of corrections are shown for the
$Z\tau\tau$ form factor $d^Z_\tau(m_Z^2)$ in Fig. 4. They
are of second order in the SM couplings. \\
It may be instructive to discuss how these diagrams come about in specific
models -- provided (\ref{11}) applies. In models with Higgs bosons
$\varphi$ of indefinite CP parity,
diagrams Figs. 3 a,b result from the vertex functions in Figs. 2 b,a,
respectively,
when the $\varphi$ propagator is shrunk to a point, which is a reasonable
approximation if $m_{\varphi}\gg m_Z$. In these models
only the local $ZZ\tau\tau$ of (\ref{12})
and the scalar-pseudoscalar interactions of (\ref{15}) are
generated  (to Born approximation) in this limit. Hence
Figs. 3a,b, which generate the EDM and WDM form factors
as discussed above,
are, in fact, in these models one-loop effects, whereas Figs. 4 appear at
two loops and are presumably less important.\\
There are also models, for instance leptoquark models, where also the local
dipole
interactions  of (\ref{12}) are generated (to one-loop aproximation)
in the  limit (\ref{11}). Then imaginary parts of the EDM and WDM form factors
in the kinematic region $s\ll \Lambda^2$ are due to Figs. 3,4, which
are two-loop effects in this case
and hence expected to be small. \\
In summary: both types of corrections 1) and 2)
can lead to imaginary parts of the form factors $d^{\gamma,Z}_\tau
(m_Z^2)$. Moreover, these corrections make it in general necessary
to \underbar{renormalize} the coupling parameters
of the effective Lagrangian (\ref{12}):

\[\tilde d_\tau^{\gamma(0)}\to \tilde d_\tau^{\gamma(ren.)}\]
\[\tilde d_\tau^{Z(0)}\to \tilde d_\tau^{Z(ren.)}\]
The parameters $\tilde d_\tau^{\gamma,Z(ren.)}$
are then \underbar{real, finite}
quantities, which we can call the renormalized EDM and WDM coupling
constants of the effective Lagrangian.
There is a computable functional connection between
the directly measurable quantities, for instance the form factors
$d_\tau^{\gamma,Z}(m_Z^2)$ and the renormalized coupling parameters
of ${\cal L}_{eff}$:
\bear\label{16}
{\rm measurable\ quantity}\ &=&\ {\rm function\ of}\ (\tilde
d_\tau^{Z(ren.)},
\tilde d_\tau^{\gamma(ren.)}, \tilde h_{if}^{(ren.)},\nonumber\\
&&..., {\rm renormalized\ SM\ parameters)}.\ear
Such relations for $d_\tau^{\gamma,Z}(m_Z^2)$ generalize then
(\ref{14}) and give also imaginary parts to the form
factors.

The advantage of the effective Lagrangian approach is its
model-indepen\-dence. It allows us to relate
anomalous effects in various reactions to a small set of parameters
if we restrict ourselves to operators of dimension smaller
than or equal to some fixed number, e.g. six, in ${\cal L}_{\rm eff}$.
Yet using these couplings to leading order, which is
often  sufficient for a phenomenological analysis, the dependence
of ``new physics" effects on kinematic invariants is absent by construction.
In the form
factor approach, on the other hand, we have to introduce for
each reaction a set of independent form factors which
are a priori unknown functions of the relevant kinematic
variables. But form factors are more directly related to the
experimental observables. Fortunately,
in the case at hand the EDM and WDM form
factors are constant for fixed c.m. energy.
Thus a good strategy could be to have
experimentalists present their results in terms of form factors
and have theorists analyse the implications for the couplings
of the effective Lagrangian and/or of specific models beyond the SM.

At this point we can return to the question of infrared
divergences in the form factors. Due to these divergences,
the form factors at $m_Z^2$ are, strictly speaking,
\underbar{not} directly measurable quantities.
Thus, if one wants to include QED radiative corrections
in a systematic way one either has to go to the effective
Lagrangian approach anyway or one has to work with form factors
defined at the theory level before radiative corrections. It is
clear that the latter approach would be rather similar to the
effective Lagrangian one. We may also note that above we
treated the $Z$ boson as an on-shell particle, which, of course,
it never is. A more rigorous definition of the form factors in
(\ref{8}) could be given with the help of the partial wave
decomposition of the amplitude for the reaction $e^+e^-\to
\tau^+\tau^-$. But this still would leave us with the problem
of infrared divergences.

Finally let us discuss
the reaction $e^+e^-\to \tau^+\tau^-$ away from the Z pole.
We shall confine ourselves to
a discussion of CP-violating contributions. (The question of whether or not
the use of AMM and WMM form factors still
provides a gauge-invariant description of CP-invariant
chirality-flipping effects in the final state requires a  detailed
consideration within a  specific model.)
Rather than
performing a completely model-independent analysis of this amplitude
by means of a  Lorentz covariant decomposition, it is more useful
to analyse this process either in terms of effective
Lagrangians or  in terms of specific models of CP violation.
We briefly sketch the latter approach.  Consider extensions of the
Standard Model with the same gauge boson content as the SM,
and which predict, apart from the Kobayashi-Maskawa phase,
also  CP-violating "non universal" interactions of the
Higgs boson type with couplings to fermions being proportional to
the mass of a fermion  involved. Such couplings induce
 much larger dipole form factors  for heavy fermions as compared to light
flavours. It is  primarily the search for such interactions
which makes CP studies in tau physics so interesting.
Neglecting the electron mass
then in these models CP violation in the above amplitude arises
to one-loop approximation  via induced  electric and weak dipole form
factors in the $\gamma \tau^+\tau^-$ and
$Z \tau^+\tau^-$ vertex functions,
respectively.
A posteriori no strong-coupling CP phenomena
are expected in the case at hand. Hence the one-loop approximation
is sufficient. It is therefore a valid
and quite general procedure
to parameterize CP-violating effects in $e^+e^-\to \tau^+\tau^-$ in terms
of $d^{\gamma}_{\tau}(s)$ and $d^{Z}_{\tau}(s)$ as was done in
\cite{6}. It depends on the specific model
whether these form factors have, for given c.m. energy
$\sqrt s$, also imaginary parts. As discussed above
in extensions of the SM with neutral Higgs bosons of undefined CP parity
\cite{18}
the EDM form factor ${\rm Im}d^{\gamma}_{\tau}(s) \neq 0$ for
$s > 4 m^2_{\tau}$, whereas for instance in leptoquark models
$d^{\gamma}_{\tau}(s)$ will have a sizeable imaginary part only above the
$t{\bar t}$ threshold {\cite{19}}. Depending on the
c.m. energy the imaginary parts of
the form factors can become larger in magnitude than the real parts.
Future experiments should therefore search, as was done in \cite{11},
both for nonzero   ${\rm Re}d^{\gamma,Z}_{\tau}(s)$ and
${\rm Im}d^{\gamma,Z}_{\tau}(s)$.

\section{An ``Optimal'' Strategy for Searching for CP-Violation
in $Z\to b\bar bX$}

In \cite{15} a search for CP-violation in $Z\to b\bar bX$
was suggested. This seems very worthwhile since the experimental
results for the rate $Z\to b\bar bX$ deviate considerably
from
the SM prediction (cf. e.g. \cite{21}). The problem with which
experimentalists are confronted is to select
$Z\to b\bar bX$ decays from
all hadronic decays of the $Z$ and to make a CP analysis. Typically
one has a parameter $\xi$ which gives information if the hadronic
event is a $Z\to b\bar bX$ decay. This parameter $\xi$
could be the measured length $L$ between the primary and the
secondary vertex of the event divided by the width $\sigma$ of
the vertex distribution
\be\label{17}
\xi=L/\sigma.\ee
Let $w(\xi)$ be the probability that the event is a $b$-event
if $\xi$ is measured, and\\
$1-w(\xi)$ the probability that it is
a hadronic event coming from the production of $u, d, s, c$ quarks,
denoted collectively by $q$ in the following. The probability $w(\xi)$
is usually well known from Monte Carlo studies. Furthermore, let
$\rho(\xi)$ be the normalized density of events on the $\xi$-axis
\be\label{18}
\int d\xi\rho(\xi)=1.\ee

Consider  now the $Z$-decay distribution with respect to a set of
kinematic variables $\phi$ (typically jet variables) and $\xi$
for all the hadronic events, $q$ and $b$ taken together:
\be\label{19}
\frac{1}{\Gamma}\frac{d\Gamma(\phi,\xi)}
{d\phi d\xi}=\Bigl\{\left[1-w(\xi)\right]S^q_{\rm SM}(\phi)
+w(\xi)\left[S^b_{\rm SM}(\phi)+\hat h_bS^b_1(\phi)\right]\Bigr\}
\cdot\rho(\xi).\ee
Here we assume for simplicity that only $b$-quarks have
a CP-violating coupling to the $Z$ which can be parametrized
by a single coupling constant $\hat h_b$ (cf. \cite{15}).
The quantity $S^q_{\rm SM}(\phi)$ describes the SM distribution
in the kinematic variables $\phi$ for $u, d, s, c$ quarks, $S^b_{\rm SM}
(\phi)$ for the $b$ quarks and $S^b_1(\phi)$ is the CP-odd
interference term between the SM and CP-odd amplitudes for the $b$
quarks. Here and in the following terms of order $\hat h_b^2$ are
neglected.

The problem is now to devise an ``optimal'' strategy to test for the
presence of the $\hat h_b$ term in (\ref{19}). The solution
to this type of problem is well known \cite{22}. The optimal
observable is
\be\label{20}
O=\frac{w(\xi)S^b_1(\phi)}{(1-w(\xi))S^q_{\rm SM}(\phi)
+w(\xi)S^b_{\rm SM}(\phi)}.\ee
To make things simple, let us consider the case where the
SM distributions $S^q_{\rm SM}(\phi)$ and
$S^b_{\rm SM}(\phi)$ are equal:
\be\label{21}
S^q_{\rm SM}(\phi)=S^b_{\rm SM}(\phi).\ee
Usually this is a good approximation, valid to within a few
percent.
We get then
\be\label{22}
O=w(\xi)\frac{S_1^b(\phi)}{S^b_{\rm SM}(\phi)},\ee
\bear\label{23}
<O>&=&\hat h_b\int d\xi\rho(\xi)w^2(\xi)\int d\phi\frac{[S^b_1(\phi)]^2}
{S^b_{\rm SM}(\phi)},\nonumber\\
<O^2>&=&\int d\xi\rho(\xi)w^2(\xi)\int d\phi\frac{[S^b_1(\phi)]^2}
{S^b_{\rm SM}(\phi)}\ear
and for the 1 s.d. error of an (ideal) measurement of $\hat h_b$ via
$O$:
\be\label{24}
(\delta\hat h_b)_{opt}^2=\frac{<O^2>}{N(<O>/\hat h_b)^2}=\left[
\int d\phi\frac{[S^b_1(\phi)]^2}
{S^b_{\rm SM}(\phi)}\right]^{-1}N^{-1}_{b,eff}.\ee
Here
\be\label{25}
N_{b,eff}:=N\int d\xi \rho(\xi)w^2(\xi)\ee
and $N$ is the total number of $Z$ decays considered.
Since we have $0\leq w(\xi)\leq1$, the total number of $b$-events
\be\label{26}
N_b=N\int d\xi\rho(\xi)w(\xi)\ee
is always bigger or equal $N_{b,eff}$:
\be\label{27}
N_b\geq N_{b,eff}.\ee
Thus, the error $(\delta \hat h_b)^2_{opt}$ in (\ref{24}) is
larger or equal to the error one would have
in an ideal experiment, where
all $b$-events could be selected unambiguously, which would
correspond to the replacement $N_{b,eff}\to N_b$ in (\ref{24}).

On the other hand, the procedure of making cuts on $b$-events
in a region where $w(\xi)$ is large, say for
\be\label{28}
\xi_1\leq\xi\leq\xi_2\ee
and using as observable the optimal one, disregarding $\xi$:
\be\label{29}
\tilde O=\frac{S^b_1(\phi)}
{S^b_{\rm SM}(\phi)},\ee
produces an error larger or equal to the one in (\ref{24}).
Indeed, in this case we get for the number of $b$-events
within the cuts (\ref{28})
\be\label{30}
N_{b,cut}:=N\int^{\xi_2}_{\xi_1}d\xi\rho(\xi)w(\xi)\ee
and for the expectation values of $\tilde O,\tilde O^2$ and
for $(\delta \hat h_b)^2_{cut}$:
\be\label{31}
<\tilde O>_{cut}=\hat h_b\frac{N_{b,cut}}{N}\int d\phi
\frac{[S^b_1(\phi)]^2}
{S^b_{\rm SM}(\phi)},\ee
\be\label{32}
<\tilde O^2>_{cut}=\int^{\xi_2}_{\xi_1}d\xi\rho(\xi)\int d\phi
\frac{[S^b_1(\phi)]^2}
{S^b_{\rm SM}(\phi)},\ee
\be\label{33}
(\delta\hat h_b)^2_{cut}=\left[\int d\phi\frac{[S^b_1(\phi)]^2}
{S^b_{\rm SM}(\phi)}\right]^{-1}\tilde N_{b,cut}^{-1},\ee
where
\be\label{34}
\tilde N_{b, cut}:=N^2_{b,cut}\left[N\cdot\int^{\xi_2}_
{\xi_1}d\xi\rho(\xi)\right]^{-1}.\ee

Using the Cauchy-Schwartz inequality we find easily
\[N^2_{b,cut}\leq N^2\int^{\xi_2}_
{\xi_1}d\xi'\rho(\xi')\int^{\xi_2}_
{\xi_1}d\xi\rho(\xi)w^2(\xi),\]
\bear\label{35}
N^2_{b,cut}\left[ N\int^{\xi_2}_
{\xi_1}d\xi\rho(\xi)\right]^{-1}&\leq &N\int^{\xi_2}_
{\xi_1}d\xi\rho(\xi)w^2(\xi)\nonumber\\
&\leq &N\int
d\xi\rho(\xi)w^2(\xi).\ear
Thus
\be\label{36}
\tilde N_{b,cut}\leq N_{b,eff},\ee
 \be\label{37}
(\delta\hat h_b)^2_{cut}\geq(\delta\hat h_b)^2_{opt},\ee
q.e.d.

\section{Conclusions}

We hope that this note will be useful for experimentalists and
interesting for theorists.

\section*{Acknowledgements}

We would like to thank P. Haberl, J. von Krogh, P. Overmann, A. Stahl,
N. Wermes, and
M. Wunsch for many discussions concerning CP-tests. Special thanks are due to
A. Stahl and N. Wermes for suggesting to write this note and
discussions on the manuscript,  and P. Overmann
for his help with the figures. One of the authors
(O.N.) would like to thank V. Telegdi for discussions and correspondence
concerning the anapole moment.

\section*{Figure Captions}
\begin{description}
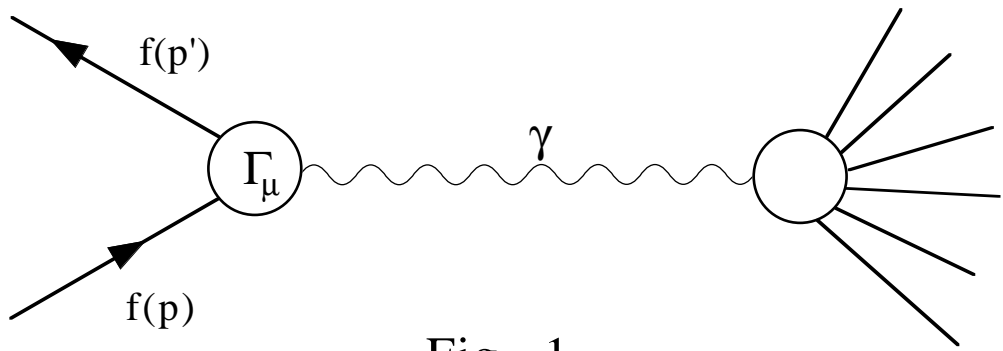
\item{Fig. 1}
Photon exchange diagram which contains  the $ff\gamma$-vertex
(1).
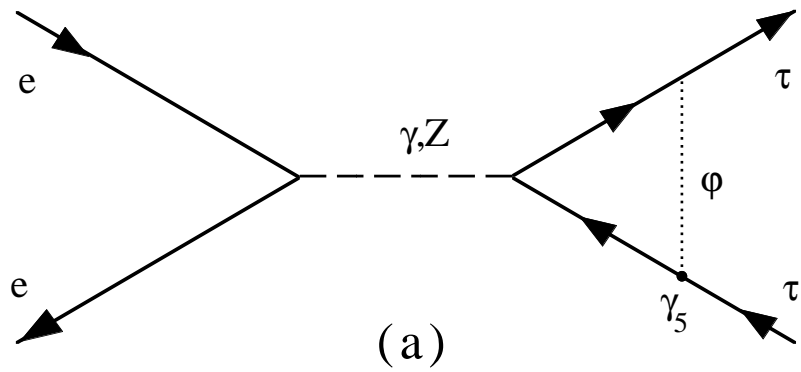
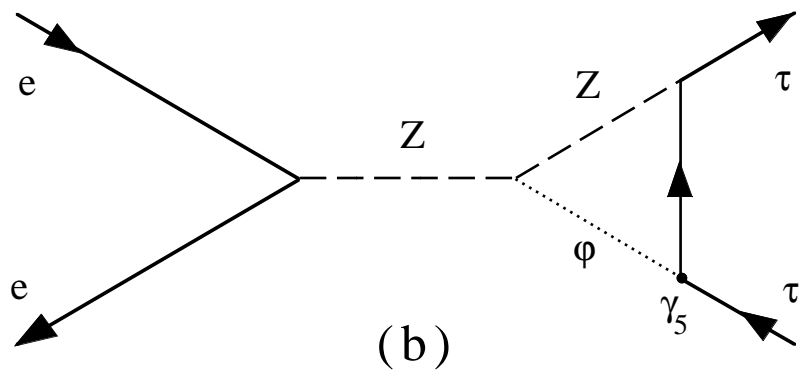
\item{Fig. 2} CP-violating contributions to $e^+ e^-\to \tau^+\tau^-$
(for $m_e$=0) in models with Higgs bosons $\varphi$ of undefined CP parity.
Diagrams with permuted vertices are not drawn.
\item{Fig. 3} Diagrams which involve the CP-odd $ZZ\tau\tau$
coupling (a) and the CP-odd 4-point
couplings (b) which
give a contribution to the dipole form factor $d^Z_\tau(m_Z^2)$.
\item{Fig. 4} Radiative corrections to the $Z\tau\tau$ dipole coupling
in the SM.
\end{description}

\end{document}